\documentclass[10pt]{wlscirep}
\usepackage{graphicx, color}
\newcommand{\Tr}{\mathrm{Tr}}

\title{Enhanced Information Exclusion Relations}

\author[1, 2]{Yunlong Xiao}
\author[3, 1 *]{Naihuan Jing}
\author[2]{Xianqing Li-Jost}
\affil[1]{School of Mathematics, South China University of Technology, Guangzhou 510640, China}
\affil[2]{Max Planck Institute for Mathematics in the Sciences, Leipzig 04103, Germany}
\affil[3]{Department of Mathematics, North Carolina State University, Raleigh, NC 27695, USA}

\affil[*]{Corresponding author: jing@ncsu.edu}

\begin{abstract}
In Hall's reformulation of the uncertainty principle, the entropic uncertainty relation occupies a core position and
provides the first nontrivial bound for the information exclusion principle. Based upon recent developments on the uncertainty relation,
we present new bounds for the information exclusion relation using majorization theory and combinatoric techniques, which reveal further characteristic properties of the overlap matrix between the measurements.
\end{abstract}
\begin{document}

\flushbottom
\maketitle
%
%
\thispagestyle{empty}

\section*{Introduction}

Mutual information is a measurement of correlations and plays a central role in communication theory \cite{Cover, Holevo, Yuen} . While the entropy describes uncertainties of measurements \cite{Wehner, Bosyk, Bialynicki, Deutsch, Maassen} , mutual information quantifies bits of gained information. Furthermore, information is a more natural quantifier than entropy except in applications like transmission over quantum channels\cite{Dupuis} . The sum of information corresponding to measurements of position and momentum is bounded by the quantity $\log2\Delta X\Delta P_{X}/\hbar$ for a quantum system with uncertainties for complementary observables $\Delta X$ and $\Delta P_{X}$, and this is
equivalent to the Heisenberg uncertainty principle \cite{Heisenberg} . Both the uncertainty relation and information exclusion relation \cite{Hall, HallM, Gheorghiu} have been be used to study the complementarity of obervables such as position and momentum. The standard deviation has also been
employed to quantify uncertainties, and it has been recognized later that the entropy seems more suitable in studying certain aspects of uncertainties.

As one of the well-known entropic uncertainty relations, Maassen and Uffink's formulation \cite{Maassen} states that
\begin{align}\label{e:MU}
H(M_{1})+H(M_{2})\geqslant-\log c_{max},
\end{align}
where $H(M_k)=H(M_k, \rho)=-\sum_jp_j^k\log_2p_j^k$ with $p_j^k=\langle u_j^k|\rho|u_j^k\rangle$ $(k=1, 2; j=1, 2, \ldots, d)$ for a given density matrix $\rho$ of dimension $d$,
and $c_{max}=\max\limits_{i_{1}, i_{2}}c(u^{1}_{i_{1}}, u^{2}_{i_{2}})$, and $c(u^{1}_{i_{1}}, u^{2}_{i_{2}})=\mid\langle u^{1}_{i_{1}}|u^{2}_{i_{2}}\rangle\mid^{2}$ for two orthonormal bases $M_{1}=\{|u^{1}_{i_{1}}\rangle\}$ and $M_{2}=\{|u^{2}_{i_{2}}\rangle\}$ of
$d$-dimensional Hilbert space $\mathcal{H}$.

Hall \cite{Hall} generalized Eq.(\ref{e:MU}) to give the first bound of the {\it Information Exclusion Relation} on accessible information about a quantum system represented by an ensemble of states. Let $M_{1}$ and $M_{2}$ be as above on system $A$, and let $B$ be another classical register (which may be related to $A$), then
\begin{align}\label{e:H}
I(M_{1}: B)+I(M_{2}: B)\leqslant r_{H},
\end{align}
where $r_{H}=\log_{2}(d^{2}c_{max})$ and $I(M_{i}: B)=H(M_{i})-H(M_{i}| B)$ is the {\it mutual information} \cite{Reza} corresponding to the measurement $M_{i}$ on system $A$. Here $H(M_i|B)$ is the conditional entropy relative to the
subsystem $B$. Moreover, if system $B$ is quantum memory, then
$H(M_i|B)=H(\rho_{M_{i}B})-H(\rho_{B})$ with $\rho_{M_{i}B}=(\mathcal{M}_{i}\otimes I)(\rho_{AB})$,
while $\mathcal{M}_{i}(\cdot)=\sum_{k_{i}}|u^{i}_{k_{i}}\rangle\langle u^{i}_{k_{i}}|(\cdot)|u^{i}_{i_{k}}\rangle\langle u^{i}_{k_{i}}|$.
Eq. (\ref{e:H}) depicts that it is impossible to probe the register $B$ to reach complete information about observables $M_{1}$ and $M_{2}$ if the maximal overlap $c_{max}$ between measurements is small. Unlike the entropic uncertainty relations, the
bound $r_{H}$ is far from being tight. Grudka {\it et al.} \cite{Grudka} conjectured a stronger information exclusion relation based on numerical evidence (proved analytically in some special cases)
\begin{align}\label{e:G}
I(M_{1}: B)+I(M_{2}: B)\leqslant r_{G},
\end{align}
where $r_{G}=\log_{2}\left(d\cdot[\sum\limits_{d~largest} c(u^{1}_{i_{1}}, u^{2}_{i_{2}})]\right)$. As the sum runs over the $d$ largest $c(u^1_{i}, u^2_{j})$,
we get $r_{G}\leqslant r_{H}$, so Eq. (\ref{e:G}) is an improvement of Eq. (\ref{e:G}). Recently
Coles and Piani \cite{Coles} obtained a new information exclusion relation stronger than Eq. (\ref{e:G}), which can also be strengthened to the case of quantum memory \cite{Berta}
\begin{align}\label{e:C}
I(M_{1}: B)+I(M_{2}: B)\leqslant r_{CP}+H(A|B),
\end{align}
where
$r_{CP}=\min\{r_{CP}(M_{1},M_{2}), r_{CP}(M_{2},M_{1})\}$, $r_{CP}(M_{1},M_{2})=\log_{2}\left(d\sum\limits_{i_{1}}\max\limits_{i_{2}}c(u^{1}_{i_{1}}, u^{2}_{i_{2}})\right)$,
and $H(A|B)=H(\rho_{AB})-H(\rho_{B})$ is the conditional von Neumann entropy with $H(\varrho)=-\Tr(\varrho\log_{2}\varrho)$ the von Neumann entropy, while $\rho_{B}$ represents the reduced state of the quantum state $\rho_{AB}$ on subsystem $B$. It is clear that $r_{CP}\leqslant r_{G}$.

As pointed out in Ref. \cite{Hall} , the general information exclusion principle should have the form
\begin{align}
\sum\limits_{m=1}^{N}I(M_{m}:B)\leqslant r(M_{1}, M_{2}, \ldots, M_{N}, B),
\end{align}
for observables $M_{1}, M_{2}, \ldots, M_{N}$, where $r(M_{1}, M_{2}, \ldots, M_{N}, B)$ is a nontrivial quantum bound.  Such a quantum
bound is recently given by Zhang {\it et al.} \cite{Zhang} for the information exclusion principle of multi-measurements in the presence of quantum memory. However, almost all available bounds are not tight even for the case of two observables.

Our goal in this paper is to
give a general approach for the information and exclusion principle using new bounds for two and multiple observables of quantum systems of
 any finite dimension by generalizing Coles-Piani's uncertainty relation and using majorization techniques.
 In particular, all our results can be reduced to the case without the presence of quantum memory.

The close relationship between the information exclusion relation and the uncertainty principle has
promoted mutual developments. In the applications of the uncertainty relation to the former, there
have been usually two available methods:
either through subtraction of the uncertainty relation in the presence of quantum memory or utilizing the concavity property of the entropy
together with combinatorial techniques or certain symmetry. Our second goal in this work is to
 analyze these two methods and in particular, we will
show that the second method together with a special combinatorial scheme enables us to find tighter bounds for the information exclusion principle.
The underlined reason for effectiveness is due to the special composition of the mutual information.
We will take full advantage of this phenomenon and apply a distinguished symmetry of cyclic permutations to derive new bounds, which
would have been difficult to obtain without consideration of mutual information.

We also remark that the recent result \cite{JX} for the sum of entropies is valid in the absence of quantum side information and cannot be extended to the
cases with quantum memory by simply adding the conditional entropy between the measured particle and quantum memory. To resolve this difficulty, we use a different method in this
paper to generalize the results of Ref. \cite{JX} in Lemma 1 and Theorem 2 to allow for quantum memory.

\section*{Results}

We first consider the information exclusion principle for two observables, and then generalize it to multi-observable cases. After that we will show that our information exclusion relation gives a tighter bound, and the bound not only involves the $d$ largest $c(u^{1}_{i_{1}}, u^{2}_{i_{2}})$ but also contains all the overlaps $c(u^{1}_{i_{1}}, u^{2}_{i_{2}})$ between bases of measurements.

We start with a qubit system to show our idea.
The bound offered by Coles and Piani for two measurements does not improve the previous bounds for quibit systems. To see these,
set $c_{i_{1}i_{2}}=c(u^{1}_{i_{1}}, u^{2}_{i_{2}})$ for brevity, then the unitarity of overlaps between measurements implies that
$c_{11}+c_{12}=1$, $c_{11}+c_{21}=1$, $c_{21}+c_{22}=1$ and $c_{12}+c_{21}=1$. Assuming $c_{11}>c_{12}$, then $c_{11}=c_{22}>c_{12}=c_{21}$, thus
\begin{align}\label{e:qubit}
&r_{H}=\log_{2}(d^{2}c_{max})=\log_{2}(4c_{11}),\notag\\
&r_{G}=\log_{2}(d\sum\limits_{d~largest} c(u^{1}_{i_{1}}, u^{2}_{i_{2}}))=\log_{2}(2(c_{11}+c_{22})),\notag\\
&r_{CP}=\min\{r_{CP}(M_{1},M_{2}), r_{CP}(M_{2},M_{1})\}=\log_{2}(2(c_{11}+c_{22})),
\end{align}
hence we get $r_{H}=r_{G}=r_{CP}=\log_{2}(4c_{11})$ which says that the bounds of Hall, Grudka {\it et al},  and Coles and Piani
coincide with each other in this case.

Our first result already strengthens the bound in this case. Recall the implicit bound from the tensor-product majorization
relation \cite{Rudnicki, Friedland, Puchala} is of the form
\begin{align}\label{e:hybrid}
H(M_{1}|B)+H(M_{2}|B)\geqslant-\frac{1}{2}\omega\mathfrak{B}+H(A)-2H(B),
\end{align}
where the vectors $\mathfrak{B}=(\log_{2}(\omega\cdot\mathfrak{A}_{i_{1}i_{2}}))^{\downarrow}$ and
$\mathfrak{A}_{i_1, i_2}=(c(u^{1}_{i_{1}}, u^{2}_{i})c(u^{1}_{j}, u^{2}_{i_{2}}))^{\downarrow}_{ij}$ are of size $d^2$. The symbol
${\downarrow}$ means re-arranging the components in descending order. The majorization vector bound $\omega$ for
probability tensor distributions $(p^{1}_{i_{1}}p^{2}_{i_{2}})_{i_{1}i_{2}}$ of state
$\rho$ is the $d^2$-dimensional vector
$\omega=(\Omega_1, \Omega_1-\Omega_2, \ldots, \Omega_d-\Omega_{d-1}, 0, \ldots, 0)$, where
$$\Omega_k=\max_{\rho}\sum_{|\{(i_1, i_2)\}|=k} p^{1}_{i_{1}}p^{2}_{i_{2}}.$$
The bound means that
$$(p^{1}_{i_{1}}p^{2}_{i_{2}})_{i_{1}i_{2}}\prec\omega,$$
for any density matrix $\rho$ and $\prec$ is defined by comparing the corresponding partial sums of the decreasingly rearranged vectors. Therefore $\omega$
only depends on $c_{i_{1}i_{2}}$ \cite{Rudnicki} .
We remark that the quantity $H(A)-2H(B)$ assumes a similar role as that of $H(A|B)$, which will be clarified in Theorem 2.
As for more general case of $N$ measurements, this quantity is replaced by $(N-1)H(A)-NH(B)$ in the place of $NH(A|B)$. A proof of this
relation will be given in the section of Methods. The following is our first improved information exclusion relation in a new form.

\vspace{2ex}
\noindent\textbf{Theorem 1.} For any bipartite state $\rho_{AB}$, let $M_{1}$ and $M_{2}$ be two measurements on system $A$, and $B$ the quantum memory correlated to $A$, then
\begin{align}\label{e:IEFhybrid}
I(M_{1}:B)+I(M_{2}:B)\leqslant2+\frac{1}{2}\omega\mathfrak{B}+2H(B)-H(A),
\end{align}
where $\omega$ is the majorization bound
 and $\mathfrak B$ is defined in the paragraph under Eq. (\ref{e:hybrid}).

See Methods for a proof of Theorem 1.

\begin{figure}
\centering
\includegraphics[width=0.45\textwidth]{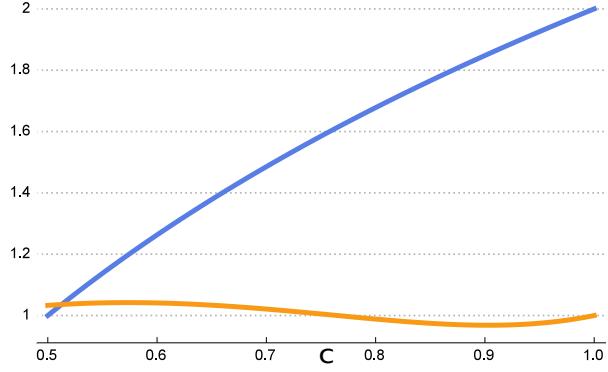}
\caption{First comparison with Hall's bound. The upper orange curve (our bound $2+\frac{1}{2}\omega\mathfrak{B}$) is tighter than
 the lower blue one (Hall's bound $r_{H}$) almost everywhere.
}
\end{figure}

\begin{figure}
\centering
\includegraphics[width=0.45\textwidth]{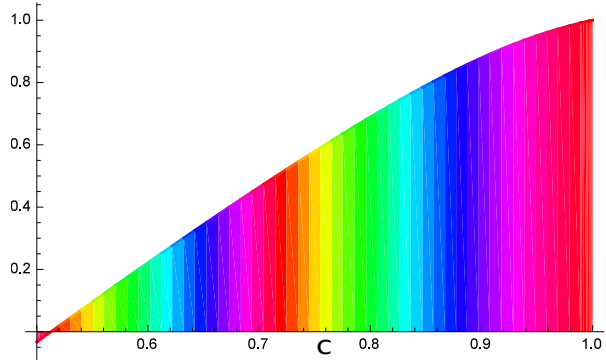}
\caption{First comparison with Hall's bound. The difference $r_{H}-2-\frac{1}{2}\omega\mathfrak{B}$ of our bound from Hall's bound $r_H$ for $a\in [0.5, 1]$ is shown.}
\end{figure}

Eq. (\ref{e:IEFhybrid}) gives an implicit bound for the information exclusion relation, and it is tighter
than $\log_{2}(4c_{max})+2H(B)-H(A)$ as our bound not only involves the maximal overlap between $M_{1}$ and $M_{2}$, but also contains the second largest element based on the
construction of the universal uncertainty relation $\omega$
\cite{Friedland, Puchala} . Majorization approach \cite{Friedland, Puchala} has been widely used in improving the lower bound of entropic uncertainty relation.
The application in the information exclusion relation offers a new aspect of the majorization method.
The new lower bound not only can be used for arbitrary nonnegative Schur-concave function \cite{Marshall}
such as R\'{e}nyi entropy and Tsallis entropy \cite{Tsallis} , but also provides insights to the relation among all the overlaps between measurements, which explains why
it offers a better bound for both entropic uncertainty relations and information exclusion relations.
We also remark that the new bound is still weaker than the one based on the optimal entropic
uncertainty relation for qubits \cite{Ghirardi} .

 As an example, we consider the measurements $M_{1}=\{(1, 0), (0, 1)\}$ and $M_{2}=\{(\sqrt{a}, e^{i\phi}\sqrt{1-a}), (\sqrt{1-a}, -e^{i\phi}\sqrt{a})\}$.
 Our bound and $\log_{2}4c_{max}$ for $\phi=\pi/2$ with respect to $a$ are shown in FIG. 1.

FIG. 1 shows that our bound for qubit is better than the previous bounds $r_{H}=r_{G}=r_{CP}$ almost everywhere.
Using symmetry we only consider $a$ in $[\frac{1}{2},1]$. The common term $2H(B)-H(A)$ is omitted in the comparison.
Further analysis of the bounds is given in FIG. 2.

Theorem 1 holds for any bipartite system and can be used for arbitrary two measurements $M_{i}$ $(i=1, 2)$. For example, consider
the qutrit state and a family of unitary matrices $U(\theta)=M(\theta)O_{3}M(\theta)^{\dag}$ \cite{Coles, Rudnicki} where
\begin{align}
M(\theta)=
\left(
\begin{array}{ccc}
  1 & 0 & 0 \\
  0 & \cos\theta & \sin\theta \\
  0 & -\sin\theta & \cos\theta \\
\end{array}
\right),\notag\\
O_{3}=
\frac{1}{\sqrt{6}}\left(
\begin{array}{ccc}
  \sqrt{2} & \sqrt{2} & \sqrt{2} \\
  \sqrt{3} & 0 & -\sqrt{3} \\
  1 & -2 & 1 \\
\end{array}
\right).
\end{align}
Upon the same matrix $U(\theta)$,
comparison between our bound $2+\frac{1}{2}\omega\mathfrak{B}$ and Coles-Piani's bound $r_{CP}$ is depicted in FIG. 3.

\begin{figure}
\centering
\includegraphics[width=0.45\textwidth]{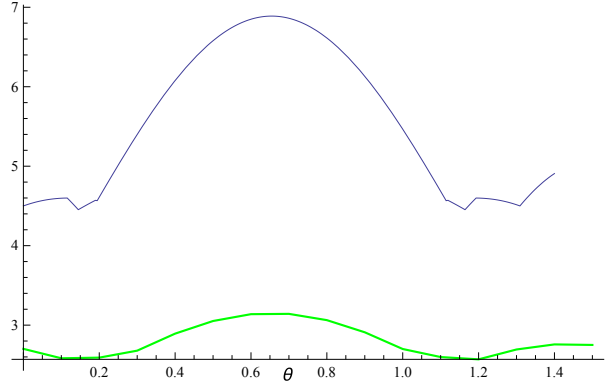}
\caption{Comparison of our bound with that of Coles and Piani.
Our bound $2+\frac{1}{2}\omega\mathfrak{B}$ (lower in green) is better than Coles-Piani's bound $r_{CP}$ (upper in purple) everywhere.
}
\end{figure}

In order to generalize the information exclusion relation to multi-measurements, we recall that the universal bound of tensor products of
two probability distribution vectors can be
computed by optimization over minors of the overlap matrix \cite{Friedland, Puchala} . More generally for the multi-tensor product
$(p^{1}_{i_{1}}p^{2}_{i_{2}}\cdots p^{N}_{i_{N}})$ corresponding to measurement $M_{m}$ on a fixed quantum state, there exists similarly
a universal upper bound $\omega$: $(p^{1}_{i_{1}}p^{2}_{i_{2}}\cdots p^{N}_{i_{N}})\prec\omega$. Then we have the following lemma, which
generalizes Eq. \eqref{e:hybrid}.

\vspace{2ex}
\noindent\textbf{Lemma 1.} For any bipartite state $\rho_{AB}$, let $M_{m}$ $(m=1, 2, \ldots, N)$  be $N$ measurements on system $A$, and $B$ the quantum memory correlated to $A$, then the following entropic uncertainty relation holds,
\begin{align}\label{e:admixture}
\sum\limits_{m=1}^{N}H(M_{m}|B)\geqslant-\frac{1}{N}\omega\mathfrak{B}+(N-1)H(A)-NH(B),
\end{align}
where $\omega$ is the $d^N$-dimensional majorization bound for the $N$ measurements $M_{m}$
and $\mathfrak B$ is the $d^N$-dimensional vector $(log(\omega\cdot\mathfrak{A}_{i_{1}, i_{2}, \ldots, i_{N}}))^{\uparrow}$
defined as follows.
For each multi-index $(i_{1}, i_{2}, \ldots, i_{N})$, the $d^{N}$-dimensional vector $\mathfrak{A}_{i_{1}, i_{2}, \ldots, i_{N}}$
has entries of the form $c(1, 2, \ldots, N)c(2, 3, \ldots, 1)\cdots c(N, 1, \ldots, N-1)$ sorted in decreasing order with respect to the indices $(i_{1}, i_{2}, \ldots, i_{N})$ where $c(1, 2, \ldots, N)=\sum\limits_{i_{2}, \ldots, i_{N-1}}\max\limits_{i_{1}}c(u^{1}_{i_{1}}, u^{2}_{i_{2}})\cdots c(u^{N-1}_{i_{N-1}}, u^{N}_{i_{N}})$ .

See Methods for a proof of Lemma 1.

We remark that the {\it admixture bound} introduced in Ref. \cite{JX} was based upon the majorization theory with the help of the action of the symmetric group,
and it was shown that the bound outperforms previous results. However, the admixture bound cannot be extended to the entropic uncertainty relations in the presence
of quantum memory for multiple measurements directly. Here we first use a new method to generalize the results of Ref. \cite{JX} to allow for the quantum side
information by mixing properties of the conditional entropy and Holevo inequality in Lemma 1. Moreover, by combining Lemma 1 with properties
of the entropy we are able to give an enhanced information exclusion relation (see Theorem 2 for details).

The following theorem is obtained by subtracting entropic uncertainty relation from the above result.

\vspace{2ex}
\noindent\textbf{Theorem 2.} For any bipartite state $\rho_{AB}$, let $M_{m}$ $(m=1, 2, \ldots, N)$  be $N$ measurements on system $A$, and $B$ the quantum memory correlated to $A$, then
\begin{align}
\sum\limits_{m=1}^{N}I(M_{m}:B)\leqslant\log_{2}d^{N}+\frac{1}{N}\omega\mathfrak{B}+NH(B)-(N-1)H(A):=r_{x},
\end{align}
where $\frac{1}{N}\omega\mathfrak{B}$ is defined in Eq. (\ref{e:admixture}).

See Methods for a proof of Theorem 2.

\begin{figure}
\centering
\includegraphics[width=0.45\textwidth]{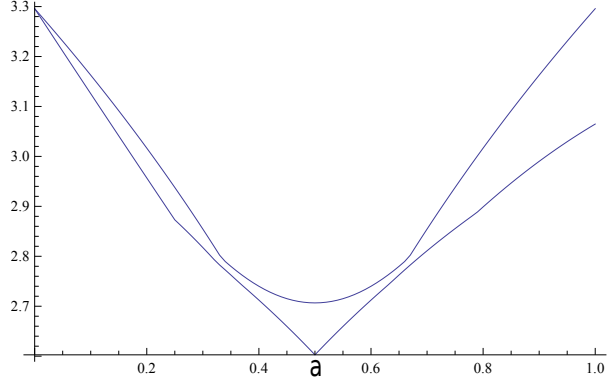}
\caption{Comparison of our bound with that of Zhang et al.
Our bound $r_{x}$ in the bottom is tighter than the top curve of Zhang's bound $\widetilde{\mathcal{U}}_{1}$.}
\end{figure}

Throughout this paper, we take $NH(B)-(N-1)H(A)$ instead of $-(N-1)H(A|B)$ as the variable that quantifies the amount of entanglement between
measured particle and quantum memory since $NH(B)-(N-1)H(A)$ can outperform $-(N-1)H(A|B)$ numerically to some extent for entropic uncertainty
relations.

Our new bound for multi-measurements offers an improvement than the bound recently given in Ref. \cite{Zhang} . Let us recall the information exclusion relation bound \cite{Zhang} for multi-measurements (state-independent):
\begin{align}
\sum\limits_{m=1}^{N}I(M_{m}:B)\leqslant\min\left\{\widetilde{\mathcal{U}}_{1}, \widetilde{\mathcal{U}}_{2}, \widetilde{\mathcal{U}}_{3}\right\}
\end{align}
with the bounds $\widetilde{\mathcal{U}}_{1}$, $\widetilde{\mathcal{U}}_{2}$ and $\widetilde{\mathcal{U}}_{3}$ are defined as follows:
\begin{align*}
\widetilde{\mathcal U}_1&=N\log_2d+NH(B)-(N-1)H(A)+\min_{(i_1\ldots i_N)\in\mathfrak S_N}
\left\{\log\max_{i_N}\{\sum_{i_2\ldots i_{N-1}}\max_{i_1}\prod_{n=1}^{N-1}c(u_{i_n}^n, u_{i_{n+1}}^{n+1})\}\right\},\\
\widetilde{\mathcal U}_2&=(N-1)\log_2d+NH(B)-(N-1)H(A)+\min_{(i_1\ldots i_N)\in\mathfrak S_N}
\left\{\log\sum_{i_2\ldots i_{N}}\max_{i_1}\prod_{n=1}^{N-1}c(u_{i_n}^n, u_{i_{n+1}}^{n+1})\right\},\\
\widetilde{\mathcal U}_3&=N\log_2d+\frac N2(2H(B)-H(A))+\frac1{|I_2|}\sum_{(k, l)\in I_2}
\left\{\min\{\log\max_{i_k}c(u_{i_k}^k, u_{i_l}^{l}), \log\max_{i_l}c(u_{i_k}^k, u_{i_l}^{l})\}\right\}.
\end{align*}
Here the first two maxima are taken over all permutations $(i_1i_2\ldots i_N): j\to i_j$, and the third is over all
possible subsets
$I_2=\{(k_1, l_1), \ldots, (k_{|I_2|}, l_{|I_2|})\}$
such that $(k_1, l_1, \ldots, k_{|I_2|}, l_{|I_2|})$
 is a $|I_2|$-permutation $1, \ldots, N$.
For example, $(12)$, $(23)$, $\ldots$, $(N-1, N)$, $(N1)$ is a $2$-permutation of $1, \ldots, N$, while $(12), (13), \ldots, (N-1, N), (N, 1)$ is an
$(N-1)$-permutation of $1, \ldots, N$. Clearly, $\widetilde{\mathcal U}_3$ is the average value of all potential two-measurement combinations.

Using the permutation symmetry, we have the following Theorem which improves the bound $\widetilde{\mathcal U_3}$.

\vspace{2ex}
\noindent\textbf{Theorem 3.} Let $\rho_{AB}$ be the bipartite density matrix with measurements $M_{m}$ $(m=1, 2, \ldots, N)$
on the system $A$ with a quantum memory $B$ as in Theorem 2, then
\begin{align}
\sum\limits_{m=1}^{N}I(M_{m}:B)\leqslant& N\log_{2}d+\frac{N}{2}(2H(B)-H(A))\notag\\
+&\frac1{|I_L|}\sum_{(k_{1}, k_{2}, \ldots, k_{L})\in I_L}
\left\{\min_{(k_{1}, k_{2}, \ldots, k_{L})}\{\log\max_{i_{L}}\sum_{k_2, \ldots, k_{L-1}}\max_{k_{1}}
\prod\limits_{n=1}^{L-1}c(u_{k_{n}}^n, u_{k_{n+1}}^{n+1})\}\right\}:=r_{opt},
\end{align}
where the minimum is over all $L$-permutations of $1, \ldots, N$ for $L=2, \ldots, N$.

In the above we have explained that the bound $\widetilde{\mathcal U_3}$ is obtained by taking the minimum over
all possible 2-permutations of $1, 2, \ldots, N$, naturally our new bound $r_{opt}$ in Theorem 3 is sharper than
$\widetilde{\mathcal U_3}$ as we have
considered all possible multi-permutations of $1, 2, \ldots, N$.

Now we compare $\widetilde{\mathcal{U}}_{1}$ with $r_{x}$. As an example in three-dimensional space, one chooses three measurements as follows \cite{Liu} :
$$u^{1}_{1}=(1, 0, 0), u^{1}_{2}=(0, 1, 0), u^{1}_{3}=(0, 0, 1);$$
$$u^{2}_{1}=(\frac{1}{\sqrt{2}}, 0, -\frac{1}{\sqrt{2}}), u^{2}_{2}=(0, 1, 0), u^{2}_{3}=(\frac{1}{\sqrt{2}}, 0, \frac{1}{\sqrt{2}});$$
$$u^{3}_{1}=(\sqrt{a}, e^{i\phi}\sqrt{1-a}, 0), u^{3}_{2}=(\sqrt{1-a}, -e^{i\phi}\sqrt{a}, 0), u^{3}_{3}=(0, 0, 1).$$

FIG 4 shows the comparison when $a$ changes and $\phi=\pi/2$, where it is clear that $r_x$ is better than
$\widetilde{\mathcal U_1}$.

The relationship between $r_{opt}$ and $r_{x}$ is sketched in FIG. 5. In this case
$r_{x}$ is better than $r_{opt}$ for three measurements of dimension three, therefore $\min\{r_{opt}, r_{x}\}=\min\{r_{x}\}$.
Rigorous proof that $r_{x}$ is always better than $r_{opt}$ is nontrivial, since all the possible combinations of measurements less than $N$
must be considered.

\begin{figure}
\centering
\includegraphics[width=0.45\textwidth]{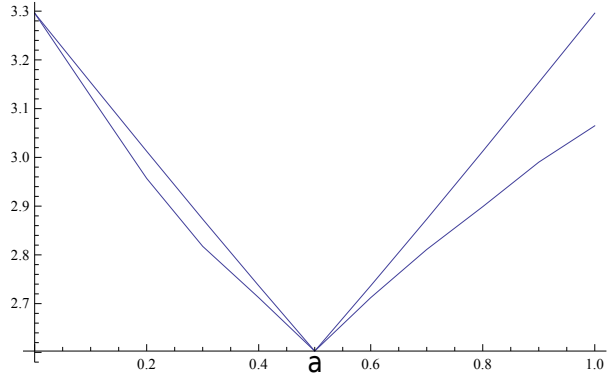}
\caption{Comparison of our two bounds via combinatorial and majorization methods: the top curve is $r_{opt}$ (combinatorial),
while the lower curve is $r_x$ (majorization).}
\end{figure}

On the other hand, we can give a bound better than $\widetilde{\mathcal{U}}_{2}$. Recall that the concavity has been utilized in the formation of $\widetilde{\mathcal{U}}_{2}$, together with all possible combinations we will get following lemma (in order to simplify the process, we first consider cases of three measurements, then generalize it to multiple measurements).

\vspace{2ex}
\noindent\textbf{Lemma 2.} For any bipartite state $\rho_{AB}$, let $M_{1}, M_2, M_3$ be three measurements on system $A$
in the presence of quantum memory $B$,
then
\begin{align}\label{e:log}
\sum\limits_{m=1}^{3}I(M_{m}:B)\leqslant-2H(A|B)+\sum_{cyclic~perm}\log_{2}\left[\sum\limits_{i_{3}}\left(\sum\limits_{i_{2}}\max\limits_{i_{1}}
c(u^{1}_{i_{1}}, u^{2}_{i_{2}})c(u^{2}_{i_{2}}, u^{3}_{i_{3}})\right)^{\frac{1}{3}}\right],
\end{align}
where the sum is over the three cyclic permutations of $1,2,3$.

See Methods for a proof of Lemma 2.

Observe that the right hand side of Eq. (\ref{e:log}) adds the sum of three terms $$\frac{1}{3}\sum\limits_{i_{3}}p^{3}_{i_{3}}\log_{2}[\sum\limits_{i_{2}}\max\limits_{i_{1}}c(1, 2, 3)], \ \ \frac{1}{3}\sum\limits_{i_{1}}p^{1}_{i_{1}}\log_{2}[\sum\limits_{i_{3}}\max\limits_{i_{2}}c(2, 3, 1)], \ \
\frac{1}{3}\sum\limits_{i_{2}}p^{2}_{i_{2}}\log_{2}[\sum\limits_{i_{1}}\max\limits_{i_{3}}c(3, 1, 2)].$$
Naturally, we can also add $\frac{1}{2}\sum\limits_{i_{3}}p^{3}_{i_{3}}\log_{2}[\sum\limits_{i_{2}}\max\limits_{i_{1}}c(1, 2, 3)]$ and
$\frac{1}{2}\sum\limits_{i_{1}}p^{1}_{i_{1}}\log_{2}[\sum\limits_{i_{3}}\max\limits_{i_{2}}c(2, 3, 1)]$. By the same method, consider all possible combination and denote the minimal as $r_{3y}$. Similar for $N$-measurements, set the minimal bound under the concavity of logarithm function as $r_{Ny}$, moreover let $r_{y}=\min\limits_{m}\{r_{my}\}$ $(1\leqslant m\leqslant N)$, hence $r_{y}\leqslant\widetilde{\mathcal{U}}_{2}$, finally we get

\vspace{2ex}
\noindent\textbf{Theorem 4.} For any bipartite state $\rho_{AB}$, let $M_{m}$ $(m=1, 2, \ldots, N)$  be $N$ measurements on system $A$, and $B$
 the quantum memory correlated to $A$, then
\begin{align}
\sum\limits_{m=1}^{N}I(M_{m}:B)\leqslant\min\{r_{x}, r_{y}\}
\end{align}
with $\frac{1}{N}\omega\mathfrak{B}$ the same in Eq. (\ref{e:admixture}). Since $\min\{r_{x}, r_{y}\}\leqslant\min\{\widetilde{\mathcal{U}}_{1}, \widetilde{\mathcal{U}}_{2}, \widetilde{\mathcal{U}}_{opt}\}$ and all figures have shown our newly construct bound $\min\{r_{x}, r_{y}\}$ is tighter. Noted that there is no clear relation between $r_{x}$ and $r_{y}$, while the bound $r_{y}$ cannot be obtained
by simply subtracting the bound of entropic uncertainty relations in the presence of quantum memory. Moreover, if
$r_{y}$ outperforms $r_{x}$, then we can utilize $r_{y}$ to achieve new bound for entropic uncertainty
relations stronger than $-\frac{1}{N}\omega\mathfrak{B}$.

\section*{Conclusions}

We have derived new bounds of the information exclusion relation for multi-measurements in the presence of quantum memory.
The bounds are shown to be tighter than recently available bounds by detailed illustrations.
Our bound is obtained by utilizing the concavity of the entropy function. The procedure
has taken into account of all possible permutations of the measurements, thus offers a significant improvement than previous results
which had only considered part of 2-permutations or combinations. Moreover, we have shown that majorization of the probability distributions for multi-measurements offers better bounds. In summary, we have formulated a systematic method of finding tighter bounds by combining the symmetry
principle with majorization theory, all of which have been made easier in the context of mutual information. We remark that the new bounds can be easily computed by numerical computation.

\section*{Methods}

\noindent\textbf{Proof of Theorem 1.} Recall that the quantum relative entropy $D(\rho||\sigma)=\Tr(\rho\log_2\rho)-\Tr(\rho\log_2\sigma)$
satisfies that $D(\rho||\sigma)\geqslant D(\tau\rho||\tau\sigma)\geqslant 0$ under any quantum channel $\tau$. Denote by $\rho_{AB}\to\rho_{M_1B}$ the quantum channel
$\rho_{AB}\rightarrow
\sum_i |u_i^1\rangle\langle u_i^1|\rho_{AB}|u_i^1\rangle\langle u_i^1|$, which is also
$\rho_{M_1B}=\sum_i|u_i^1\rangle\langle u_i^1|\otimes\Tr_A(\rho_{AB}|u_i^1\rangle\langle u_i^1|)$. Note that both $M^i=\{|u_j^i\rangle\} (i=1, 2)$ are  measurements on system $A$, we have that
for a bipartite state $\rho_{AB}$
\begin{align*}
H(M_{1}|B)-H(A|B)&=H(\rho_{M_1B})-H(\rho_{AB})=\Tr(\rho_{AB}\log_2\rho_{AB})-\Tr(\rho_{M_1B}\log_2\rho_{M_1B})\\
&=D(\rho_{AB}\|\sum\limits_{i_{1}}|u^{1}_{i_{1}}\rangle\langle u^{1}_{i_{1}}|\otimes \Tr_A (\rho_{AB}|u^{1}_{i_{1}}\rangle\langle u^1_{i_1}|))\notag.
\end{align*}
Note that $\Tr_B\Tr_A (\rho_{AB}|u^{1}_{i}\rangle\langle u^1_i|)=p_i^1$,
the probability distribution of the reduced state $\rho_A$ under the measurement $M_1$,
so $\sigma_{B_i}=\Tr_A (\rho_{AB}|u^{1}_{i}\rangle\langle u^1_{i}|)/p_i^1$ is a density matrix on the system $B$.
Then the last expression can be written as
\begin{align*}
&D(\rho_{AB}\|\sum\limits_{i_{1}}p_{i_1}^1|u^{1}_{i_{1}}\rangle\langle u^{1}_{i_{1}}|\otimes \sigma_{B_{i_1}})\\
\geqslant& D(\rho_{M_2B}\|\sum\limits_{i_{1}, i_2}p_{i_1}^1C_{i_1i_2}|u^{2}_{i_{2}}\rangle\langle u^{2}_{i_{2}}|\otimes \sigma_{B_{i_1}})).
\end{align*}
If system $B$ is a classical register, then we can obtain
\begin{align}
H(M_{1})+H(M_{2})\geqslant
H(A)-\sum\limits_{i_{2}}p^{2}_{i_{2}}\log\sum\limits_{i_{1}}p^{1}_{i_{1}}c(u^{1}_{i_{1}}, u^{2}_{i_{2}}),
\end{align}
by swapping the indices $i_{1}$ and $i_{2}$, we get that
\begin{align}
H(M_{2})+H(M_{1})\geqslant
H(A)-\sum\limits_{i_{1}}p^{1}_{i_{1}}\log\sum\limits_{i_{2}}p^{2}_{i_{2}}c(u^{2}_{i_{2}}, u^{1}_{i_{1}}).
\end{align}
Their combination implies that
\begin{align}
H(M_{1})+H(M_{2})\geqslant
H(A)-\frac{1}{2}\left(\sum\limits_{i_{2}}p^{2}_{i_{2}}\log\sum\limits_{i_{1}}p^{1}_{i_{1}}c(u^{1}_{i_{1}}, u^{2}_{i_{2}})
+\sum\limits_{i_{1}}p^{1}_{i_{1}}\log\sum\limits_{i_{2}}p^{2}_{i_{2}}c(u^{2}_{i_{2}}, u^{1}_{i_{1}})\right),
\end{align}
thus it follows from Ref. \cite{NC} that
\begin{align}
H(M_{1}|B)+H(M_{2}|B)\geqslant
H(A)-2H(B)-\frac{1}{2}\left(\sum\limits_{i_{2}}p^{2}_{i_{2}}\log\sum\limits_{i_{1}}p^{1}_{i_{1}}c(u^{1}_{i_{1}}, u^{2}_{i_{2}})
+\sum\limits_{i_{1}}p^{1}_{i_{1}}\log\sum\limits_{i_{2}}p^{2}_{i_{2}}c(u^{2}_{i_{2}}, u^{1}_{i_{1}})\right),
\end{align}
hence
\begin{align}
I(M_{1}|B)+I(M_{2}|B)&= H(M_{1})+H(M_{2})-(H(M_{1}|B)+H(M_{2}|B))\notag\\
&\leqslant H(M_{1})+H(M_{2})+\frac{1}{2}\left(\sum\limits_{i_{2}}p^{2}_{i_{2}}\log\sum\limits_{i_{1}}p^{1}_{i_{1}}c(u^{1}_{i_{1}}, u^{2}_{i_{2}})
+\sum\limits_{i_{1}}p^{1}_{i_{1}}\log\sum\limits_{i_{2}}p^{2}_{i_{2}}c(u^{2}_{i_{2}}, u^{1}_{i_{1}})\right)+2H(B)-H(A)\notag\\
&\leqslant H(M_{1})+H(M_{2})+\frac{1}{2}\sum\limits_{i_{1},i_{2}}p^{1}_{i_{1}}p^{2}_{i_{2}}\log_{2}\left(\sum\limits_{i,j}
p^{1}_{i}p^{2}_{j}c(u^{1}_{i}, u^{2}_{i_{2}})c(u^{2}_{j}, u^{1}_{i_1})\right)+2H(B)-H(A)\notag\\
&\leqslant 2+\frac{1}{2}\omega\mathfrak{B}+2H(B)-H(A),
\end{align}
where the last inequality has used $H(M_{i})\leqslant\log_{2}d$ $(i=1, 2)$ and the vector $\mathfrak{B}$ of length $d^{2}$, whose entries $\mathfrak{B}_{i_{1}i_{2}}=\log_{2}(\omega\cdot\mathfrak{A}_{i_{1}i_{2}})$ are arranged in decreasing order with respect to $(i_{1}, i_{2})$. Here the vector $\mathfrak A$ is defined by $\mathfrak{A}_{i_{1}i_{2}}=c(u^{1}_{i}, u^{2}_{i_{2}})c(u^{2}_{j}, u^{1}_{i_1})$ for each $(i_{1}, i_{2})$ and also sorted in decreasing order. Note that
the extra term $2H(B)-H(A)$ is another quantity appearing on the right-hand side that describes the amount of entanglement between
the measured particle and quantum memory besides $-H(A|B)$.

We now derive the {\it information exclusion relation} for qubits in the form of
$I(M_{1}:B)+I(M_{2}:B)\leqslant2+\frac{1}{2}\omega\mathfrak{B}+2H(B)-H(A)$, and this completes the proof.
\vspace{2ex}

\noindent\textbf{Proof of Lemma 1.} Suppose we are given $N$ measurements $M_1, \ldots, M_N$ with orthonormal bases $\{|u_{i_j}^j\rangle\}$. To simplify presentation we denote that
\begin{align*}
c_{i_1, \ldots, i_N}^{1,\ldots, N}=c(u^{1}_{i_{1}}, u^{2}_{i_{2}})c(u^{2}_{i_{2}}, u^{3}_{i_{3}})\cdots c(u^{N-1}_{i_{N-1}}, u^{N}_{i_{N}}).
\end{align*}
Then we have that \cite{Liu}
\begin{align}
(1-N)H(A)+\sum\limits_{m=1}^{N} H(M_{m})
\geqslant&-\Tr(\rho \log \sum\limits_{i_{1}, i_{2}, \ldots, i_{N}} p^{1}_{i_{1}}c_{i_1, \ldots, i_N}^{1,\ldots, N}|u^N_{i_N}\rangle\langle u^N_{i_N}|)
=-\sum\limits_{i_{N}}p^{N}_{i_{N}}\log \sum\limits_{i_{1}, i_{2}, \ldots, i_{N-1}} p^{1}_{i_{1}}c_{i_1, \ldots, i_N}^{1,\ldots, N}.
\end{align}
Then consider the action of the cyclic group of $N$ permutations on indices $1, 2, \cdots, N$, and
taking the average gives the following inequality:
\begin{align}
\sum\limits_{m=1}^{N}H(M_{m})\geqslant-\frac{1}{N}\omega\mathfrak{B}+(N-1)H(A),
\end{align}
where the notations are the same as appeared in Eq. (\ref{e:admixture}). 
Thus it follows from Ref. \cite{NC} that
\begin{align}
\sum\limits_{m=1}^{N}H(M_{m}|B)\geqslant-\frac{1}{N}\omega\mathfrak{B}+(N-1)H(A)-NH(B).
\end{align}
The proof is finished.

\vspace{2ex}

\noindent\textbf{Proof of Theorem 2.} Similar to the proof of Theorem 1, due to $I(M_{m}:B)=H(M_{m})-H(M_{m}|B)$, thus we get
\begin{align}
\sum\limits_{m=1}^{N}I(M_{m}:B)&= \sum\limits_{m=1}^{N}H(M_{m})-\sum\limits_{m=1}^{N}H(M_{m}|B)\notag\\
&\leqslant \sum\limits_{m=1}^{N}H(M_{m})+\frac{1}{N}\omega\mathfrak{B}+NH(B)-(N-1)H(A)\notag\\
&\leqslant \log_{2}d^{N}+\frac{1}{N}\omega\mathfrak{B}+NH(B)-(N-1)H(A),
\end{align}
with the product $\frac{1}{N}\omega\mathfrak{B}$ the same in Eq. (\ref{e:admixture}).

\vspace{2ex}

\noindent\textbf{Proof of Lemma 2.} First recall that for $\sum\limits_{i_{3}}p^{3}_{i_{3}}\log_{2}[\sum\limits_{i_{2}}\max\limits_{i_{1}}c(u^{1}_{i_{1}}, u^{2}_{i_{2}})c(u^{2}_{i_{2}}, u^{3}_{i_{3}})]$ we have
\begin{align*}
&\quad H(M_3)+\sum\limits_{i_{3}}p^{3}_{i_{3}}\log_{2}[\sum\limits_{i_{2}}\max\limits_{i_{1}}c(u^{1}_{i_{1}}, u^{2}_{i_{2}})c(u^{2}_{i_{2}}, u^{3}_{i_{3}})]\\
&=\sum\limits_{i_{3}}p^{3}_{i_{3}}\log_{2}\frac{1}{p^{3}_{i_{3}}}+
\sum\limits_{i_{3}}p^{3}_{i_{3}}\log_{2}[\sum\limits_{i_{2}}\max\limits_{i_{1}}c(u^{1}_{i_{1}}, u^{2}_{i_{2}})c(u^{2}_{i_{2}}, u^{3}_{i_{3}})]\\
&\leqslant \log_{2}\sum\limits_{i_{3}}[\sum\limits_{i_{2}}\max\limits_{i_{1}}c(u^{1}_{i_{1}}, u^{2}_{i_{2}})c(u^{2}_{i_{2}}, u^{3}_{i_{3}})],
\end{align*}
where we have used concavity of $\log$. By the same method we then get
\begin{align}
\sum\limits_{m=1}^{3}I(M_{m}:B)\leqslant&\sum\limits_{m=1}^{3}H(M_{m})+3H(B)-2H(A)+
\frac{1}{3}\sum\limits_{i_{3}}p^{3}_{i_{3}}\log_{2}[\sum\limits_{i_{2}}\max\limits_{i_{1}}c(1, 2, 3)]\notag\\
&+\frac{1}{3}\sum\limits_{i_{2}}p^{2}_{i_{2}}\log_{2}[\sum\limits_{i_{1}}\max\limits_{i_{3}}c(3, 1, 2)]
+\frac{1}{3}\sum\limits_{i_{1}}p^{1}_{i_{1}}\log_{2}[\sum\limits_{i_{3}}\max\limits_{i_{2}}c(2, 3, 1)]\notag\\
=&\sum\limits_{m=1}^{3}H(M_{m})+3H(B)-2H(A)+
\sum\limits_{i_{3}}p^{3}_{i_{3}}\log_{2}[\sum\limits_{i_{2}}\max\limits_{i_{1}}c(1, 2, 3)]^{\frac{1}{3}}\notag\\
&+\sum\limits_{i_{2}}p^{2}_{i_{2}}\log_{2}[\sum\limits_{i_{1}}\max\limits_{i_{3}}c(3, 1, 2)]^{\frac{1}{3}}
+\sum\limits_{i_{1}}p^{1}_{i_{1}}\log_{2}[\sum\limits_{i_{3}}\max\limits_{i_{2}}c(2, 3, 1)]^{\frac{1}{3}}\notag\\
\leqslant&3H(B)-2H(A)
+\log_{2}\sum\limits_{i_{3}}[\sum\limits_{i_{2}}\max\limits_{i_{1}}c(1, 2, 3)]^{\frac{1}{3}}
+\log_{2}\sum\limits_{i_{2}}[\sum\limits_{i_{1}}\max\limits_{i_{3}}c(3, 1, 2)]^{\frac{1}{3}}
+\log_{2}\sum\limits_{i_{1}}[\sum\limits_{i_{3}}\max\limits_{i_{2}}c(2, 3, 1)]^{\frac{1}{3}}\notag\\
=&3H(B)-2H(A)\notag\\
&+\log_{2}\left\{\left[\sum\limits_{i_{3}}\left(\sum\limits_{i_{2}}\max\limits_{i_{1}}c(1, 2, 3)\right)^{\frac{1}{3}}\right]\left[\sum\limits_{i_{2}}\left(\sum\limits_{i_{1}}\max\limits_{i_{3}}c(3, 1, 2)\right)^{\frac{1}{3}}\right]
\left[\sum\limits_{i_{1}}\left(\sum\limits_{i_{3}}\max\limits_{i_{2}}c(2, 3, 1)\right)^{\frac{1}{3}}\right]\right\},
\end{align}
with $c(1, 2, 3)$, $c(2, 3, 1)$ and $c(3, 1, 2)$ the same as in Eq. (\ref{e:log}) and this completes the proof.

\section*{Acknowledgments}

We would like to thank Tao Li for help on computer diagrams.
This work is supported by
National Natural Science Foundation of China (grant nos. 11271138, 11531004), China Scholarship Council and Simons Foundation grant no. 198129.

\section*{Author contributions statement}

Y. X. and N. J. analyzed and wrote the manuscript. They reviewed the paper together with X. L.-J.

\section*{Additional information}

\textbf{Competing financial interests:} The authors declare no competing financial interests.
\vspace{1ex}

\leftline{\textbf{How to cite this article:} Xiao, Y., Jing, N. and Li-Jost, X. Enhanced Information Exclusion Relations. Sci. Rep. (2016), Article no. ---}

\end{document}